\documentclass[%
 aip,
 amsmath,amssymb,
 reprint,%
]{revtex4-1}

\usepackage{graphicx}
\usepackage{dcolumn}
\usepackage{bm}
\usepackage{amsmath}

\usepackage[utf8]{inputenc}
\usepackage[T1]{fontenc}
\usepackage{mathptmx}

\begin{document}

\preprint{AIP/123-QED}

\title[Laser Stabilization Yb Pol. Enh. FM Spectro]{Laser stabilisation to neutral Yb in a discharge with polarization enhanced frequency modulation spectroscopy}

\author{Valdis Bl\={u}ms}
\author{Jordan Scarabel}
\author{Kenji Shimizu}
\author{Moji Ghadimi}
\author{Steven C. Connell}
\author{Sylvi H\"{a}ndel}
\author{Benjamin G. Norton}
\author{Elizabeth M. Bridge}
\author{David Kielpinski}
\affiliation{ 
Centre for Quantum Dynamics, Griffith University, Brisbane, Queensland 4111, Australia}
\author{Mirko Lobino}
\affiliation{ 
Centre for Quantum Dynamics, Griffith University, Brisbane, Queensland 4111, Australia}
\affiliation{Queensland Micro- and Nanotechnology Centre, Griffith University, Brisbane, QLD 4111 Australia}

 \author{Erik W. Streed}
    \email{e.streed@griffith.edu.au}
\affiliation{ 
Centre for Quantum Dynamics, Griffith University, Brisbane, Queensland 4111, Australia}

    \affiliation{Institute for Glycomics, Griffith University, Gold Coast, QLD 4222 Australia}

\date{\today}

\begin{abstract}
Isotope selective optical excitation of atoms is important for experiments with neutral atoms, metrology, and work with trapped ions, including quantum information processing. Polarisation-enhanced absorption spectroscopy is used to frequency stabilise a tunable external cavity laser diode system at $398.9$~nm for isotope selective photoionization of neutral Yb atoms. This spectroscopy technique is used to measure isotope resolved dispersive features from transitions within a see through configuration Ytterbium hollow-cathode discharge lamp. This Doppler-free dichroic polarisation spectroscopy is realised by retro-reflecting a laser beam through the discharge and analysing the polarisation dependent absorption with balanced detection. The spectroscopy signal is recovered using lock-in detection of frequency modulation induced by current modulation of the external cavity laser diode. Here, we show the use of polarisation-enhanced absorption spectroscopy for isotope selectively loading of $^{171}\mbox{Yb}^{+}$, $^{172}\mbox{Yb}^{+}$, or $^{174}\mbox{Yb}^{+}$ ions into an RF Paul trap.
\end{abstract}

\maketitle

\section{\label{sec:Intro}Introduction}
The $398.9$ nm $^{1}$S$_{0}$$\leftrightarrow$$^{1}$P$_{1}$ transition is commonly used in both Yb and Yb$^{+}$ atomic physics experiments due to its large dipole moment and accessibility with low-cost laser diodes. For neutral Yb this transitions allows for faster cooling and a broader velocity capture range of atoms than the narrower $555.8$~nm $^{1}$S$_{0}$$\leftrightarrow$$^{3}$P$_{1}$ transition, eliminating the need for a Zeeman slower \cite{Rapol2004}, while still being suitable for sub-Doppler cooling \cite{Kostylev2014}. For generating Yb$^{+}$ ions the $398.9$~nm transition provides a well resolved isotope selective excitation step in the ionisation process, which is frequently completed by continuum ionisation by the Yb$^{+}$ UV cooling laser at $369.5$~nm. This approach is more economical and accessible than using an isotope-enriched Yb source combined with non-isotope selective electron impact ionization\cite{Ender1993}.

Frequency-stable lasers, tunable to within a few MHz of an atomic resonance, are now a standard tool for precision atomic physics experiments \cite{Hansch2006,Cornell2002,Takasu2003,McLoughlin2011,Rohde2010}. Common approaches to frequency stabilise a CW laser include interfering it with a stabilized optical frequency comb \cite{Hansch2006,Shao2019} as well as using a Fabry-P\'{e}rot transfer cavity \cite{McLoughlin2011,Rohde2010,Zhao1998,Mohle2013} or commerical Fizeau-interferometer-based wavemeter \cite{Saleh2015,Couturier2018} to transfer the frequency stability from one laser wavelength to another. All of these options require an existing stable frequency reference, which can result in a greater cost, greater complexity, or potentially lower robustness over directly stabilising the laser frequency to the desired atomic frequency standard.

Numerous spectroscopy techniques have been reported~\cite{Hall2006}, primarily using atomic or molecular vapor cells~\cite{Demtroeder}. Traditional vapour cells are not suitable for low vapour pressure/high boiling point elements including Ytterbium. At room temperature Yb has a vapour pressure 11 orders of magnitude lower than Caesium or Rubidium. Increasing the temperature to produce comparable pressures results in Yb reacting chemically with the glass windows of a vapor cell. Efforts have been made to design and build vapor cells for elements including Strontium and Ytterbium \cite{Bridge2009,Jayakumar2015}, however, there are no commercially available options. A reliable and cost-effective alternative to stabilize a UV laser is using atoms produced in the discharge of a hollow-cathode lamp. First demonstrated in 1970~\cite{Skolnick}, laser stabilization to a discharge has been extended to various atomic and molecular species, including neutral Yb~\cite{Kim2003}. Several spectroscopy techniques using Yb have been reported in an atomic oven \cite{Nizamani} or a discharge lamp including modulation transfer spectroscopy (MTS)~\cite{Wang2011}, which uses non-linear four-wave mixing, as well as dichroic atomic vapour laser locking (DAVLL)~\cite{Kim2003,Streed2008}. In addition, other experiments with hollow cathode lamps have utilised polarisation spectroscopy (PS) to measure sub-Doppler features, for both neutral and ionised Yb, that are stable with respect to frequency \cite{Zhu2014,Biercuk2014}. The ability to swap between probing dichroism or laser-induced birefringence in a discharge by changing the analysis optics have been reported \cite{Smeets2003}. 

Here we demonstrate the use of polarisation-enhanced absorption spectroscopy (POLEAS) \cite{Kunz2013} to probe the dichroism of a Yb transition. When compared to PS, POLEAS has a locking frequency stability $100$ times less sensitive to polarisation drift, aided by the different analysis optics, while still providing a larger signal-to-noise ratio than Doppler-free saturation absorption spectroscopy (D-F SAS)\cite{Haroche1972,Park2003}. POLEAS is also cheaper than MTS for this near UV transition, as it can be achieved without extra components for creating sidebands on a sub-$400$~nm laser beam. This is the first time this type of spectroscopy has been used on Ytterbium in a discharge lamp, and swapping from PS to POLEAS requires minimal changes to a spectroscopy system.

\section{\label{sec:PrincipleOfTechnique}Principle of the technique}

A key aspect of many high precision spectroscopy techniques is the ability to produce well-resolved, stable dispersive features suitable for incorporation into a frequency stabilisation servo loop. Many of these spectroscopy techniques modulate the laser light which results in a change of the response of the atomic system, such that the obtained signal from probing the sample can be passed through a lock-in amplifier to obtain a sharp and stable error signal feature.

In its most elementary form D-F SAS comprises a pump beam and a counter-propagating probe beam. The nonlinear interaction between the counter-propagating beams enable the tighter constraints in this spectroscopy technique, restricting atom interaction with both beams simultaneously to a particular velocity class, resulting in a Doppler-free spectroscopy signal. The probe beam changes properties (amplitude and polarisation) as it passes through a sample and gets sent to an analysis setup. The presence of the on-resonant pump beam saturates the transition and reduces the change in probe beam properties. The pump beam modulation is sent to a lock-in amplifier as a reference signal and is mixed with the signal obtained from the probe beam after the analysis setup. Due to the beams being counter propagating, assuming perfect overlap, the end signal is free from the heterogeneous first order Doppler broadening.

\begin{figure}
\resizebox{0.5\textwidth}{!}{  \includegraphics{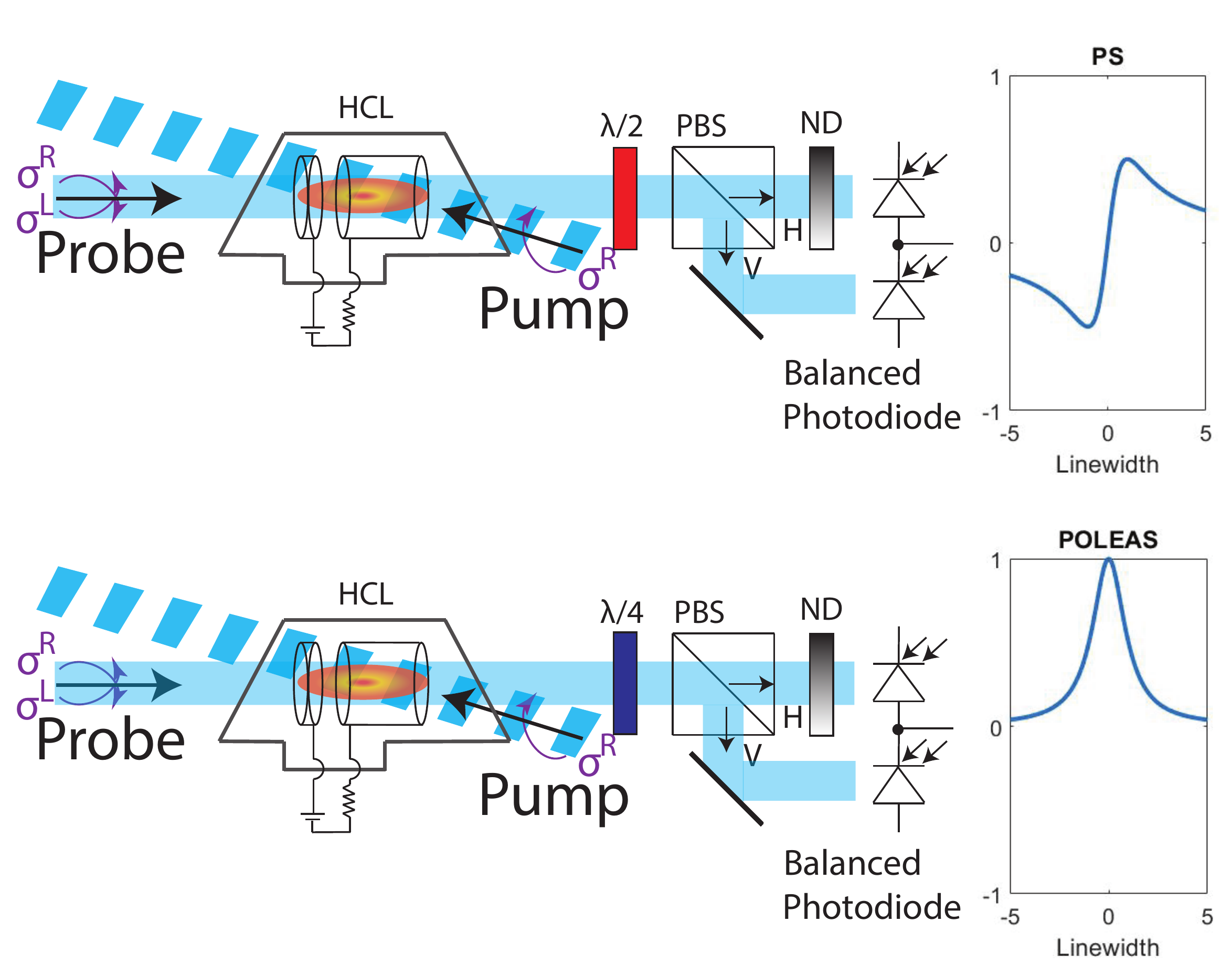}}
\caption{{\bf Top} optical setup probes laser-induced birefringence and is commonly referred to as PS. {\bf Bottom} optical setup probes dichroism and is referred to as POLEAS. The only difference between these two schematics is the wave plate used before the analysis optics, shown in red and blue. Right of each optical setup is a theoretical example of a spectroscopy feature, calculated from equations \ref{eq:complex abs_re} and \ref{eq:complex abs_im} with $s_{0}=0$ for feature clarity. For PS, $\mbox{Im}(\alpha)/\alpha_{0}$ is plotted. For POLEAS, $\mbox{Re}(\alpha)/\alpha_{0}$ is plotted. HCL, hollow-cathode lamp; PBS, polarizing beam splitter; ND, variable neutral density filter; H, horizontal polarisation; and V, vertical polarisation. }
\label{fig:polspec}
\end{figure}

Figure \ref{fig:polspec} shows a generic setup for both PS and POLEAS along with the expected output signals after lock-in detection. The setup for PS and POLEAS are similar in pump and probe beam polarisations, however, they differ in the analysis optics. A circularly polarized pump beam is sent through the sample (for convention let's call this $\sigma^R$). The counter propagating probe light is linearly polarised, which is also an equal superposition of $\sigma^R$ and $\sigma^L$. The pump beam saturates the transition for the counter propagating $\sigma^L$ light. This uses the same counter-propagating beams technique as D-F SAS and thus the Doppler broadened spread in the resonant frequency of the transition can be ignored. For a two-level atomic transition the probe light experiences a change in its properties explained by the complex absoption coefficient which, when broken into its real and imaginary parts, is \cite{Sargent}
\begin{equation}
\mbox{Re}(\alpha)=\alpha_{0}\frac{\gamma^{2}}{\gamma'^{2}+({\omega}_{0}-{\omega})^{2}}
 \label{eq:complex abs_re}
\end{equation}
and
\begin{equation}
\mbox{Im}(\alpha)=-\alpha_{0}\frac{\gamma({\omega}_{0}-{\omega})}{\gamma'^{2}+({\omega}_{0}-{\omega})^{2}}
 \label{eq:complex abs_im}
\end{equation}
where,
\begin{equation}
\gamma'=\gamma \sqrt{1+s_{0}}
 \label{eq:power_broadened}
\end{equation}
and 
\begin{equation}
{s_0} = \frac{I}{I_{\mbox{sat}}} = \frac{3I{\lambda}^{3}{\tau}}{ {\pi}hc}.
\label{eq:Isat}
\end{equation}

Here $\alpha_{0}$ is the linear resonant absorption coefficient ($s_{0}=0$), $s_{0}$ is a dimensionless on-resonance saturation parameter, $\omega$ is the laser frequency, $\omega_0$ is the transistion resonant frequency, $\gamma$ is the natural transition linewidth, and $\gamma'$ is the power-broadened linewidth. Equations \ref{eq:complex abs_re} and \ref{eq:complex abs_im} are similar to the Kramers-Kronig dispersion relations obtained when looking at the real and imaginary parts of the complex refractive index \cite{Demtroeder}. The analysis optics are set up such that for PS the real part of the complex absorption coefficient is suppressed, and for POLEAS the imaginary part is suppressed. The reason for this is as follows.

For PS, without an atomic sample, the probe beam is set to a linear polarisation such that equal amounts of both $\sigma^R$ and $\sigma^L$ light go to both detectors for analysis. In this case, the polarisation controlling $\lambda/2$ wave plate in the probe arm can be before or after the sample. When probing the sample, the change in absorption is fed into both detectors equally, and is filtered out when taking the difference between detectors. However, the $\sigma^R$ and $\sigma^L$ light experiences circular dichroism proportional to the difference in absorption coefficients $\Delta\alpha = Im(\alpha)^{+,L}-Im(\alpha)^{-,R}$. This, as well as the change in amplitude between $\sigma^R$ and $\sigma^L$ light, changes the probe beam polarisation after the lamp to be slightly elliptical. For a frequency blue shifted from resonance this elipticity sends more light towards one detector, and when the frequency is red shifted from resonance more light is sent to the other detector. This produces the dispersive spectroscopy locking signal. This setup is highly sensitive to polarisation fluctuations of the probe beam, and any fluctuation of the $\lambda/2$ wave plate from temperature, as the resulting changes to the spectroscopy signal appear similar to that caused by changes in the laser frequency.

For POLEAS, without an atomic sample, the probe beam is again set to a linear polarisation such that equal amounts of both $\sigma^R$ and $\sigma^L$ light go to both detectors. However, by placing a $\lambda/4$ wave plate after the sample all of $\sigma^R$ light can be sent to one detector while all of $\sigma^L$ light is sent to the other detector. The difference in amount of light at each detector is determined by the complex absorption coefficients for the $\sigma^R$ and $\sigma^L$ light through the sample, where $Re(\alpha)$ dominates in the end feature. The effects from $Im(\alpha)$ are suppressed as this spectroscopy is first order insensitive to polarisation fluctuations of the probe beam from induced circular dichroism, or from temperature shifts of the $\lambda/4$ wave plate. In the compressed form of POLEAS the probe beam is a retroreflected portion of the pump beam. The Doppler-free absorption profile obtained from this spectroscopy is then converted into a suitable locking signal through frequency modulation of the laser \cite{Bjorklund1980,Kobayashi1982,Lenth1983}.

\section{Ytterbium spectroscopy}\label{sec:YbSpec}
Ytterbium has a relatively straightforward lower energy level structure with optically accessible transitions which are suitable for laser cooling in both neutral and ionised Yb$^{+}$ atomic physics experiments. Ytterbium has seven naturally occurring isotopes with atomic mass units and abundances shown in table \ref{tab:yb}. Here we considered the two most abundant isotopes ($^{174}$Yb, $^{172}$Yb), as well as Yb$^{171}$ which is frequently used for quantum information processing (QIP).

\begin{table}
\caption{\label{tab:yb}Yb isotopes and their naturally occurring abundances.}
\begin{ruledtabular}
\begin{tabular}{lcd}
Yb Isotope  &\multicolumn{1}{c}{Precise Atomic mass\cite{Rana2012}}&\multicolumn{1}{c}{Abundance (\%)\cite{Laeter2006,Meija2016}}\\ \hline
 168    &                   & 0.1232(4)\\
 170    & 169.934767241(18) &   2.982(6)\\ 
 171    & 170.936331514(19) &   14.086(20)\\ 
 172    & 171.936386655(18) &   21.686(19)\\ 
 173    & 172.938216213(18) &   16.103(9)\\ 
 174    & 173.938867539(18) &   32.025(12)\\
 176    & 175.942574702(22) &   12.995(1.3)\\
\end{tabular}
\end{ruledtabular}
\end{table}

\begin{figure}
\resizebox{0.5\textwidth}{!}{%
  \includegraphics{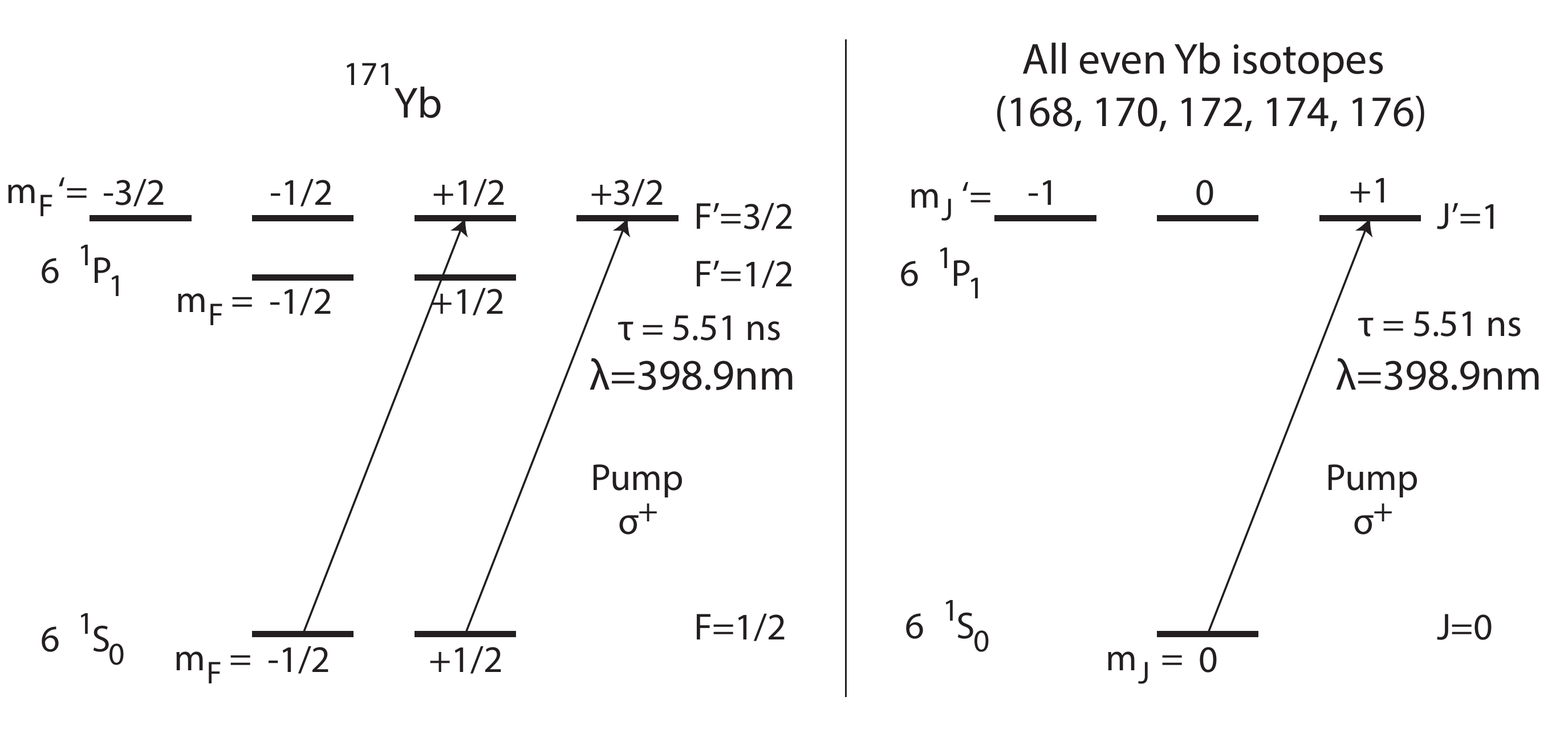}}
\caption{Relevant energy levels for the Yb isotopes producing prominent spectroscopy features suitable for laser stabilisation. Excited state lifetime $\tau= 5.51$ ns calculated from \cite{Zhao2008}. The transitions shown with arrows are saturated via the pump beam, dependent on the magnetic field direction. A more complete energy level structure can be found from \cite{Zhu2014}. For $^{171}$Yb the $^{1}$S$_{0}$ ground state $m_{F}$ levels are split by the nuclear spin while the excited $^{1}$P$_{1}$ state $m_{F}$ levels are split by the electronic and nuclear magnetic dipole interactions.
}
\label{fig:energylevels}
\end{figure}

$^{174}$Yb and $^{172}$Yb have a simpler energy level structure compared to $^{171}$Yb, because the ground-state nuclear spin is zero for even isotopes and $1/2$ for $^{171}$Yb. This nuclear spin of $1/2$, which interacts with the electron spin, splits the energy levels and results in hyperfine structure in $^{171}$Yb, see figure \ref{fig:energylevels}. For $^{171}$Yb$^{+}$ this hyperfine structure splitting of the ground state is $12.64$~GHz \cite{Fisk1997,Blatt1983}, which provides long lived microwave accessible states for QIP \cite{wang2017,Olmschenk2007}. This hyperfine splitting also provides a single F$=0$ ground state, which is ideal for optical ion clocks used in metrology \cite{Huntemann2016}. The simpler energy level structure in the even isotopes makes them useful for studying fundamental physics problems, such as observing the shadow of a single ion \cite{Streed2012} or sensing sub-attonewton forces in 3 dimensions \cite{Blums2017}, with minimal parameters and relaxes the requirements on needed equipment. This also makes them useful when setting up a new experiment for QIP applications \cite{Ghadimi2017}.

\begin{table}
\caption{\label{tab:ybfreq}Yb isotopes and their $398.9$~nm $^{1}$S$_{0}$$\leftrightarrow$$^{1}$P$_{1}$ transition frequency shifts\cite{Das2005}.}
\begin{ruledtabular}
\begin{tabular}{lcd}
Yb Isotope & Hyperfine Level& \multicolumn{1}{c}{Isotope shift (MHz)\footnote{relative to $^{174}$Yb}} \\ \hline
 168 &           &  1887.400(50) \\
 170  &          & 1192.393(66) \\
 171 &(F'=1/2)   & 1153.696(61)\\
 171 &(F'=3/2)   & 832.436(50) \\
 172 &           & 533.309(53)\\
 173 &(F'=3/2)   & 515.975(200)\\
 173 &(F'=5/2)   & 253.418(50)\\
 173 &(F'=7/2)   & 587.986(56)\\
 174 &           & 0\\
 176 &           & -509.310(50)\\
\end{tabular}
\end{ruledtabular}
\end{table}

\begin{figure}
\resizebox{0.5\textwidth}{!}{%
  \includegraphics{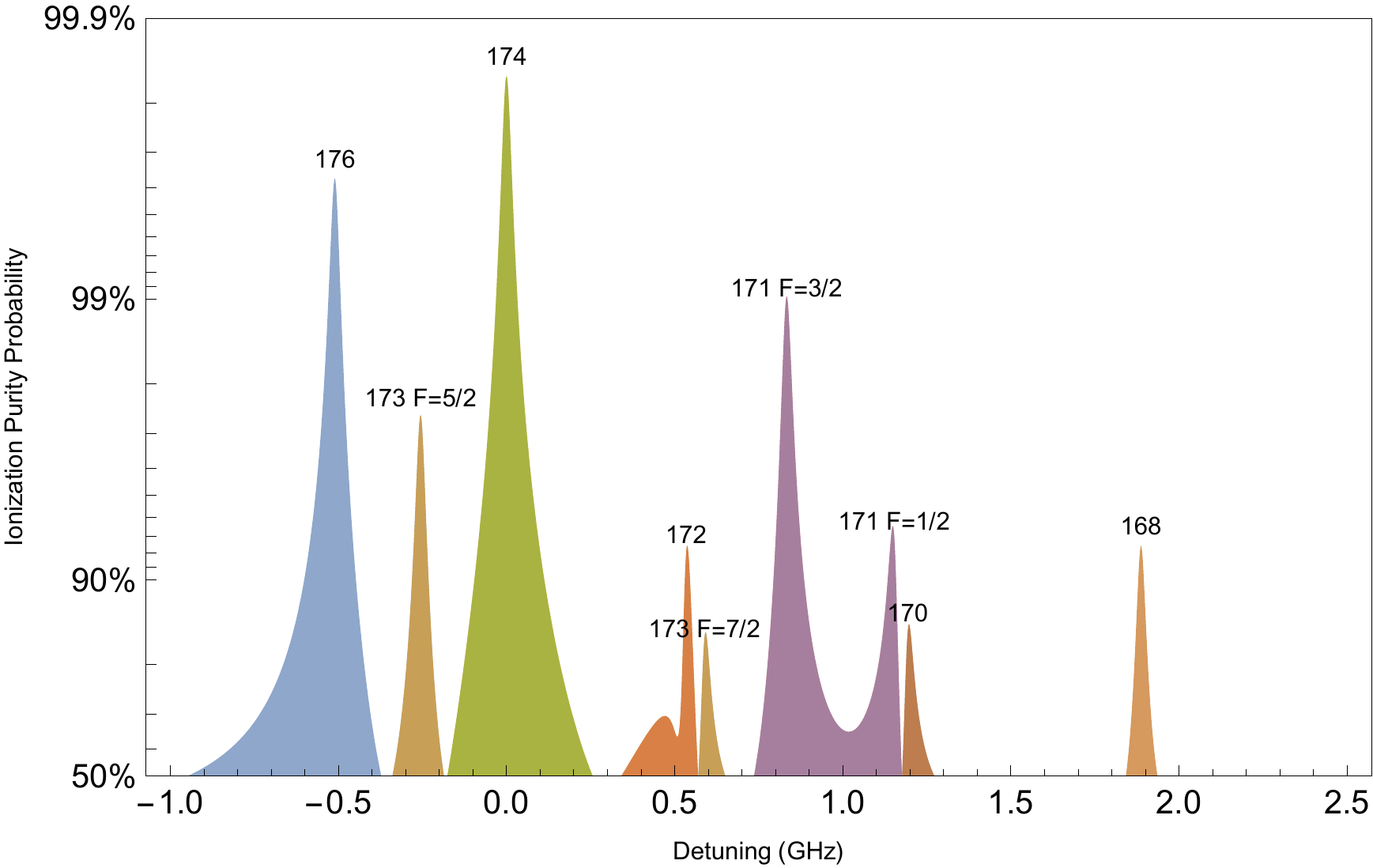}}
\caption{Isotope ionization purity rates of neutral Yb on the $398.9$~nm $^{1}$S$_{0}$$\leftrightarrow$$^{1}$P$_{1}$ transition for a natural abundance source as a function of detuning. Plot is for a collimated laser beam perpendicular to a dilute collimated atomic beam (zero Doppler broadening), in the low laser intensity limit with no optical pumping. Detunings are with respect to the transition frequency in $^{174}$Yb. Note for $^{171}$Yb, exciting the F$=3/2$ transition provides the largest probability of only exciting that isotope.}
\label{fig:PiPurity}
\end{figure}
  
Table~\ref{tab:ybfreq} and figure~\ref{fig:PiPurity} detail information on the $^{1}$S$_{0}$$\leftrightarrow$$^{1}$P$_{1}$ transition around $398.9$~nm for the different isotopes, with respect to the transition frequency in $^{174}$Yb. Table~\ref{tab:ybfreq} lists the isotope specific shift in resonant frequency and in figure~\ref{fig:PiPurity} we estimate the potential isotopic purity of ion formation from a natural abundance source based on the excitation probabilities as a function of laser detuning.

The ionisation purity $f_i$ (figure~\ref{fig:PiPurity}) is calculated for a given laser frequency $\omega$ from the relative scattering rate for a particular isotope $\Gamma_i$ vs the total scattering rate from a natural abundance distribution.

\begin{equation}
    f_i\left(\omega\right)= \frac{\Gamma_i\left(\omega\right)}{ \sum_n{ \Gamma_n\left(\omega\right)}}
\end{equation}

For isotopes 171 and 173 with nuclear spin $I\neq0$ the scattering rate is the sum across all transitions with a weighting factor $h_j$ based on the spin multiplicity.

\begin{equation}
\Gamma_i\left(\omega\right)= \sum_j h_j\frac{\gamma s_0/2}{1+s_0+  4\left(\omega_j-\omega \right)^2\gamma^{-2}}    
\end{equation}

\section{\label{sec:ExperimentalMethodResult}Experimental methods and results}
\begin{figure}
\resizebox{0.5\textwidth}{!}{%
  \includegraphics{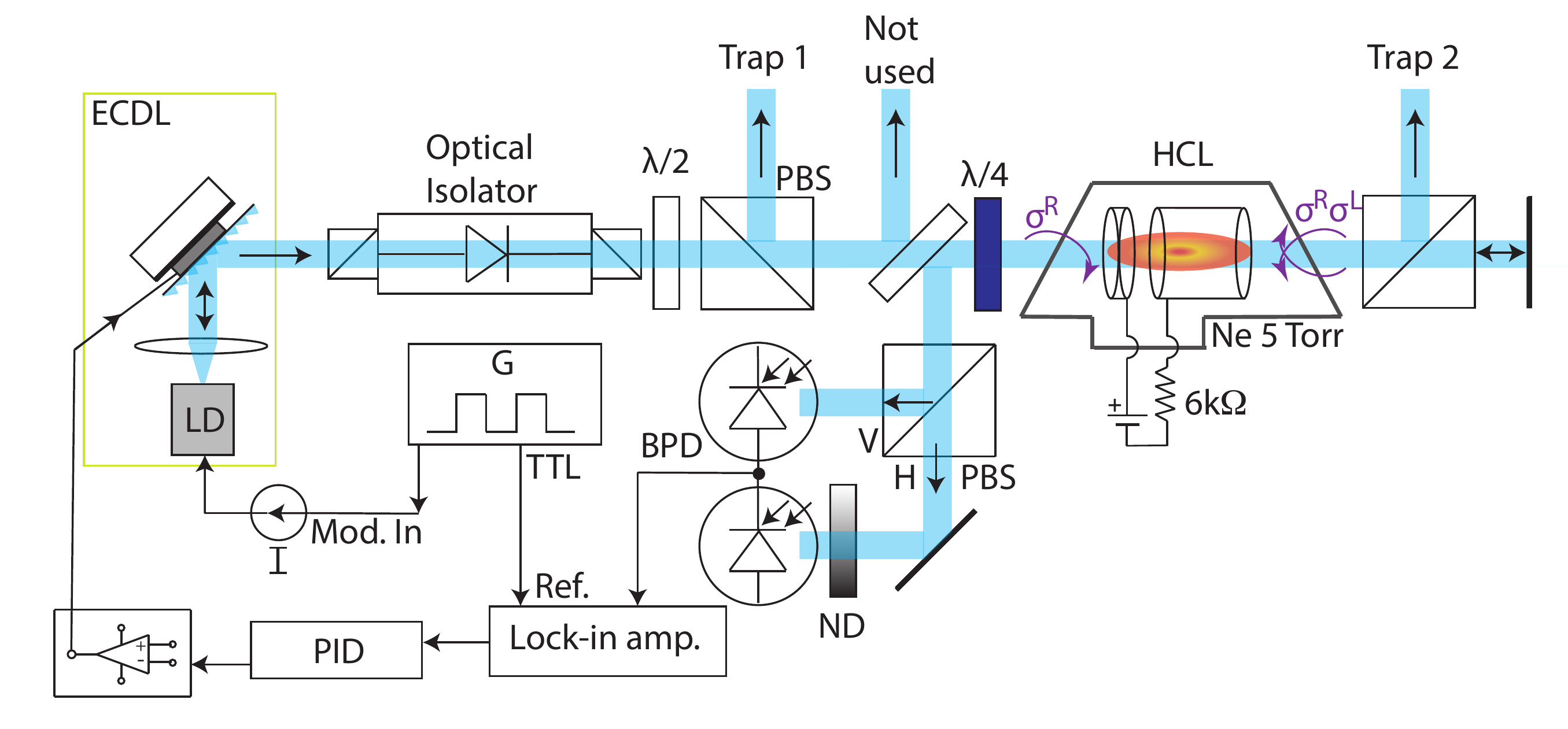}}
\caption{Optical setup for the spectroscopy using in-house external cavity diode laser (ECDL). The ECDL provides a total output power of $9$~mW after the optical isolator. Part of the light from the ECDL is picked off before the spectroscopy and fiber coupled to a $2$D RF Paul trap (trap 1). The spectroscopy is realised in a hollow-cathode lamp (HCL) and the absorption signal is recorded on a differential photodiode (BPD). After the HCL, the beam passes through a PBS where the reflected component is sent out of the setup and sent to a $3$D RF Paul trap (trap 2). Also shown are the electronics to generate the error signal, which is fed back into the laser, and the electronic circuit used for controlling the HCL. The electric potential between the HCL anode and cathode results in neutral, excited, and ionised Ne buffer gas atoms. Ne$^{+}$ ions bombard the Yb cathode, sputtering neutral Yb through the lamp \cite{Hannaford1983}. The second optical setup replaced the ECDL with a Moglabs laser system and fibre coupled $3.54(1)$~mW into the setup before the optical isolator. The third optical setup replaced the Moglabs laser with an M Squared SolsTiS plus ECD-X laser system and fibre coupled $8.26(1)$~mW into the setup before the optical isolator.
}
\label{fig:OpticalSetup}
\end{figure}

Three setups were tested for obtaining the Yb spectroscopy. The simplest one utilised an in-house assembled external-cavity diode laser (ECDL), shown in figure ~\ref{fig:OpticalSetup}. In this setup the $398.9$~nm laser system consists of an ECDL on a temperature stabilised base. The UV laser diode (Nichia Corporation LD-0397-0030-1) has a free running wavelength of $397.9$~nm, tuned to $398.9$~nm by optical feedback from a Littrow configuration $3600$ groove/mm holographic grating (Thorlabs GH$13$-$36$U). The diode is driven with a combination of direct current and current modulation averaging to 42.4~mA. This provides a total ECDL output power of $11$~mW after the grating. In order to obtain an estimate for the modulation depth used in our setup, we modulated the laser current close to DC and measured a peak-to-peak current deviation of $0.95$~mA. The beam out of the laser diode is collimated using an asphere (Thorlabs C671TME-A) providing an elliptical beam with Gaussian $1/e^{2}$ diameters of 1.89(1)~mm in the horizontal direction and 0.51(1)~mm in the vertical direction. To prevent unwanted back reflections into the diode, modulating the laser frequency or amplitude, we use an optical isolator (Conoptics 711C-1) with 80\% transmission in the forward direction and $37$~dB isolation in reverse direction. The $9$~mW beam out of the isolator gets split on a polarizing beam splitter (PBS) cube where $2$~mW of light is sent to an ion-trap experiment via a polarisation maintaining optical fiber. The remaining $7$~mW of light is turned circularly polarized, using a $\lambda/4$ wave plate, and sent through a DC hollow-cathode lamp backfilled with Ne buffer gas (Hamamastu L2783-70NE-Yb) operating at a supply current of 0.8~mA and voltage of 140~V. The choice of buffer gas species is not critical for neutral atom spectroscopy, however it does have a substantial impact on the in-lamp ion population \cite{Streed2008}.

The complete drive circuitry for this lamp is shown in figure \ref{fig:OpticalSetup}, with the DC power supply operating in current limited mode. The lamp is housed in a die-cast aluminium container with no external magnetic fields applied and designed for optical access at both sides with windows at a 10$^{\circ}$ angle to deflect back reflections. It comprises a ring cathode of length $19$~mm and a bore diameter of $3$~mm. The collimated beam exits the hollow cathode lamp and passes through a PBS, where the reflected portion of the beam is sent to a second ion-trap experiment. The transmitted portion of the beam is then retroreflected through the discharge, and $\lambda/4$ wave plate, before a fraction of that light is picked off using a glass blank and sent for analysis. That light is split on a PBS and sent to a balanced pair of photodiodes. In order to precisely balance the photocurrents we use a variable neutral density filter to balance the optical power. The difference in the detector's measured current is amplified, providing an absorption signal on a flat background. The balanced photodiode was homebuilt with a transimpedance gain of $4.6(3)$~mV/$\mu$W and a bandwidth of $20$~kHz.

\begin{figure}
\resizebox{0.5\textwidth}{!}{%
  \includegraphics{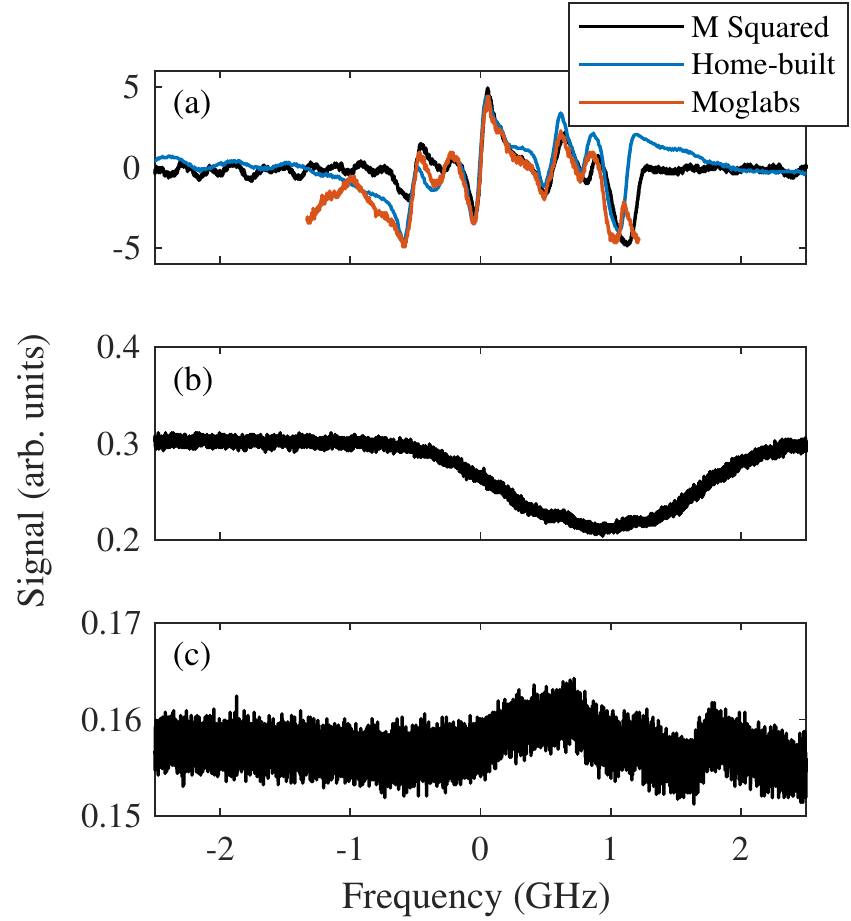}}
\caption{Laser induced spectra of neutral Yb, with the frequency axes referenced to the 174 isotope resonance. a) Differential photodiode signal after lock-in detection. The sensitivity of the lock-in amplifier was set to 1mV and the integration time on the amplifier set to 3~ms. b) Absorption signal of the discharge before lock-in detection on the differential photodiode with modulation. c) Laser induced resonance spectra of neutral Yb.}
\label{fig:specsignal}
\end{figure}

To further amplify the POLEAS signal, as well as to turn it into a zero-crossing locking signal by taking its derivative, we frequency modulated the laser through its drive current. The current modulation was done at a depth of $0.95$~mA, modulating both probe and pump beams. The laser was modulated at a frequency of $20$~kHz, with the modulation depth optimised for the largest signal-to-noise ratio without reducing feature details. Any change in the laser power from drive current modulation was filtered out of the spectroscopy signal through the balanced photodiodes. For lock-in detection we used a Stanford Research Systems SR510 lock-in amplifier, with a sensitivity of 1~mV/V and a post low-pass filter time of $0.1$~s. In figure~\ref{fig:specsignal}(a) the resulting signal out of the lock-in amplifier is shown. Changing the lamp current then changed the peak-to-peak height of the spectroscopy features. As described in \cite{Zhu2014} for the case of PS of Yb in a discharge, the size (slope) of the features increase linearly with lamp current until a tipping point is reached, measured to be $0.8$~mA for these laser settings. At higher lamp currents the absorption attenuates the probe beam reducing the signal size. The lamp current at $0.8$~mA is optimised to provide the largest spectroscopy feature for $^{174}$Yb, which coincides with the largest peak-to-peak slope for stabilising the laser frequency to. Once obtained for a particular laser system, the lamp current does not need to be changed and the feature sizes become stable.

The second setup provided similar results using a more expensive, off-the-shelf, Moglabs $399$~nm laser (Littrow Desmo) and controller (DLC102HC). This setup contained a second optical isolator with an additional $88\%$ transmission and $49$~dB isolation. The $398.9$~nm light from this laser source was fed into the system of figure~\ref{fig:OpticalSetup}, replacing the ECDL, through a polarisation maintaining fibre. The collimated-laser power of the pump beam before the HCL was $2.56(1)$ mW, with Gaussian $1/e^{2}$ diameters of $0.26(1)$~mm in the horizontal direction and $0.24(2)$~mm in the vertical direction. The diode operating current, as seen on the front display panel, was $54.78$~mA, where lasing threshold was $41.24$~mA. Current modulation amplitude for frequency modulation was $1.5$~mA. The Moglabs internal dip-switches from 1-to-16 were set to 0101 1000 0000 0000, where 0 and 1 represent the off and on states respectively. This allowed a feed-forward bias of the ECDL grating piezo voltage to diode current for a longer mode-hop free tuning range. This ratio can be measured as voltages on Ch~A and Ch~B of the Moglabs controller, with the channel outputs set to measure frequency and current respectively. This ratio of frequency/current was $7.1$ to $1$. The spectroscopy trace from this laser source is also shown in figure~\ref{fig:specsignal}(a). For this trace the best results were obtained with a lamp current of $1.4$~mA and the dual photodiodes turned off to reduce power supply noise. The spectroscopy signal was sent into the current input of the lock-in amplifier, with sensitivity set to $2$~mV/V and a low pass filter time constant of $10$~ms. 

The third, and most expensive, setup used an off the shelf M Squared lasers SolsTiS plus ECD-X system. This laser again provided similar results to the home-build ECDL, where the spectroscopy trace for this laser is also shown in figure~\ref{fig:specsignal}(a). The lock-in settings used for this laser were: sensitivity of $5$~mV/V, low pass filter time constant of $10$~ms, and modulation rate of $925$~Hz. For this laser source, figure~\ref{fig:specsignal}(b) shows the raw absorption signal observed on the balanced photodiodes.

The natural linewidth for the $398.9$~nm $^{1}$S$_{0}$$\leftrightarrow$$^{1}$P$_{1}$ transition is $\gamma=2\pi\times28.9$~MHz \cite{Zhao2008}. However, the feature sizes in Figure \ref{fig:specsignal}a are close to an order of magnitude larger ($\gamma=2\pi\times290$~MHz). The dominant mechanisms for the broadened linewidth are collisional broadening and residual Doppler broadening from the slight angle between pump and probe beams, similar to what was previously reported\cite{Zhu2014,Kimball1998}.

A neutral spectrum of Ytterbium was used for a comparative laser-frequency-to-isotope reference to the spectroscopy features. In order to obtain a neutral spectrum we produced an atomic beam of Yb in an ultra-high vacuum chamber. This was achieved by passing a high current through a resistive oven filled with natural isotope abundance Ytterbium (Alfa Aesar Yb 99.9\% REO). The emitted gas was collimated into a beam after passing through pinholes. The $398.9$~nm laser beam was sent perpendicular through the neutral beam of Yb and the laser wavelength was scanned using a signal generator sweep. When the laser was resonant with the $^{1}$S$_{0}$$\leftrightarrow$$^{1}$P$_{1}$ transition of the different isotopes, the neutral atoms in the intersection between the optical and atomic beams would fluoresce and their scattered photons were collected. From this obtained spectrum of laser-frequency dependent fluorescence, we could identify the different isotopes of Ytterbium.

\begin{figure}
\resizebox{0.5\textwidth}{!}{  \includegraphics{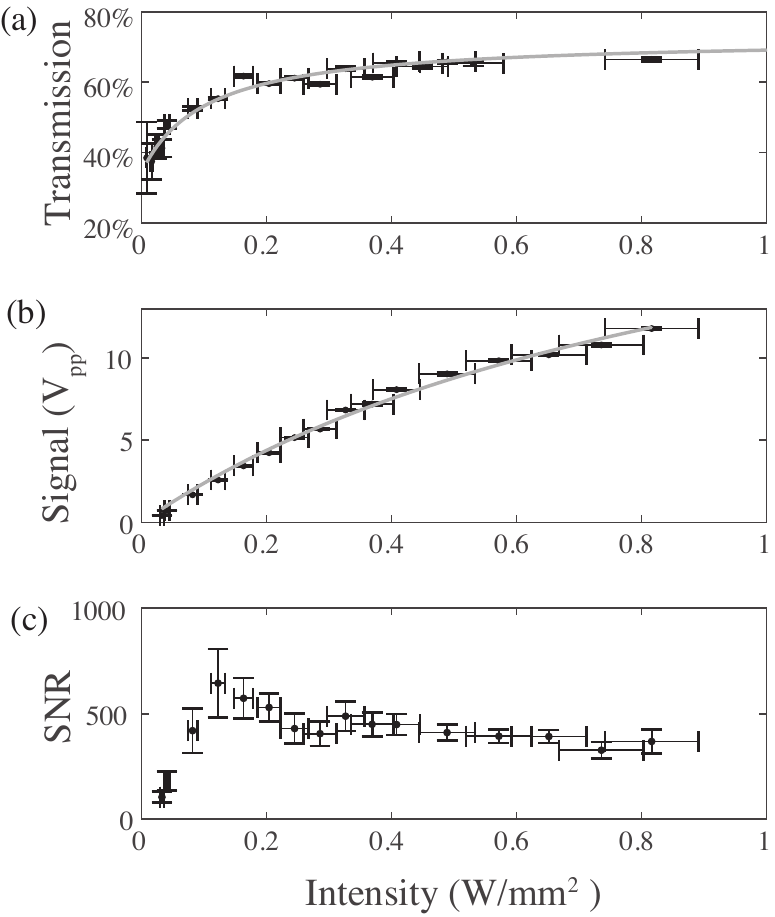}}
\caption{Circularly polarised single pass transmission and signal-to-noise ratio (SNR) of the differential signal. Intensity axis set to pump beam intensity before entering the HCL. a) Pump beam saturation curve. Line was added to guide the eye only. Line was fit to all measured data points, including those take at higher intensities (not shown) b) Spectroscopy signal size of the $^{174}$Yb spectroscopy feature. c) SNR of the prominant $^{174}$Yb spectroscopy feature, calculated as the ratio of the Signal V$_{\mbox{pp}}$ to noise V$_{\mbox{rms}}$. Above intensities of $0.1$~W/mm$^{2}$ increases in the noise dominate over signal gains, consistent with laser saturation of the transition.}
\label{fig:SNR}
\end{figure}

In order to achieve the optimal laser frequency stabilization for a given laser source, we investigated the signal-to-noise ratio of POLEAS in terms of pump beam saturation through the lamp. For an operating lamp current of 1.4~mA, the laser intensity from the high powered SolsTiS plus ECD-X source was varied while the beam size and lamp current were kept constant. In figure~\ref{fig:SNR}(a) it can be seen that for our given lamp current the absorption of a beam, through the lamp, saturates at high intensities. In figure~\ref{fig:SNR}(b) it can be seen that for low intensities the signal increases linearly with increased pump beam saturation. This is consistent with creating a larger difference in absorption coefficient between $\sigma^{+}$ and $\sigma^{-}$ transitions. The dominant noise source after lock-in detection is electronic noise from the lock-in amplifier. While the signal size continues to grow with higher input pump beam intensities, following a saturation curve, the SNR peaks at an intensity of $120(10)$ mW/mm$^{2}$ as shown in figure~\ref{fig:SNR}(c). Above this pump beam intensity the noise on the signal increases, even when using an attenuated return probe beam. 

\begin{figure}
\resizebox{0.5\textwidth}{!}{\includegraphics{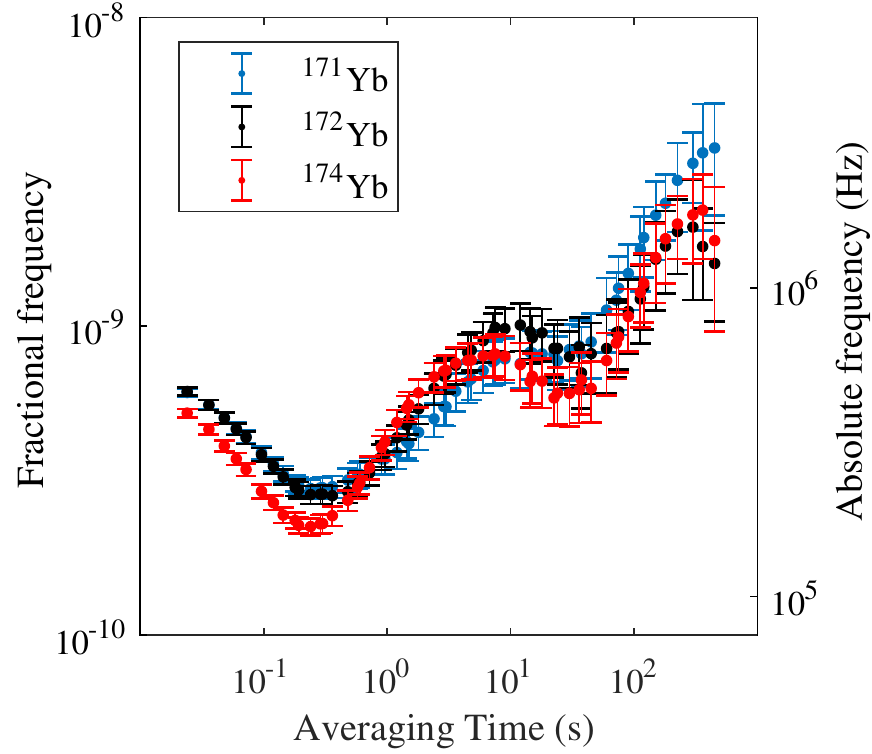}}
\caption{Frequency fluctuation of the stabilised Moglabs laser, when locked to different spectroscopy features, as a function of averaging time $\tau$. Fractional frequency fluctuations (Allan deviation $\sigma(\tau)$) scale bar on left. Absolute frequency fluctuations on right. Results are from $3~\times~1800$~s measurements. Lock-in settings were: sensitivity $200$~$\mu$V/V, low pass filter $10$~ms, and modulation rate $1.3$~kHz. }
\label{fig:AllanVariance}
\end{figure}

In order to get an estimate of the instability of the laser frequency when locked to the error feature for each of the 171, 172, and 174 Yb isotopes we measure the locked laser frequency using a precision wavemeter every $12$~ms for 1800 seconds. The wavemeter is a High Finesse WS$8$-$2$ that is calibrated to a HeNe reference laser prior to the start of each measurement. From the Allan deviation graph shown in figure~\ref{fig:AllanVariance} we observe that the laser has a similar frequency instability when locked to each of the different error features, but we cannot tell if the increased instability measured at timescales greater than 0.2 s is due to drift of the wavemeter or the laser locking approach. We have previously taken Allan deviation measurements using this wavemeter in \cite{Ghadimi2020} and from the data in this paper we expect the wavemeter fractional frequency instability to reach a minimum of around $10^{-10}$ after approximately 10 s of averaging time. Since we measure a higher fractional frequency instability than this value, we attribute the measured frequency noise to be due to instabilities deriving from the laser lock. This is most likely related to either residual uncompensated atomic density drifts in the lamp or thermally driven changes in the birefringence of optical components. 

Even with these drifts, we observe that during the $1800$~s measurement time the maximum frequency deviations were approximately $5$~MHz, which is well within the $28.9$~MHz linewidth of the atomic transition, and this is in line with our observations that the laser lock is sufficient to allow us to drive the transition and selectively photoionize the different isotopes of Ytterbium.

%
%

\begin{thebibliography}{0}%
\makeatletter
\providecommand \@ifxundefined [1]{%
 \@ifx{#1\undefined}
}%
\providecommand \@ifnum [1]{%
 \ifnum #1\expandafter \@firstoftwo
 \else \expandafter \@secondoftwo
 \fi
}%
\providecommand \@ifx [1]{%
 \ifx #1\expandafter \@firstoftwo
 \else \expandafter \@secondoftwo
 \fi
}%
\providecommand \natexlab [1]{#1}%
\providecommand \enquote  [1]{``#1''}%
\providecommand \bibnamefont  [1]{#1}%
\providecommand \bibfnamefont [1]{#1}%
\providecommand \citenamefont [1]{#1}%
\providecommand \href@noop [0]{\@secondoftwo}%
\providecommand \href [0]{\begingroup \@sanitize@url \@href}%
\providecommand \@href[1]{\@@startlink{#1}\@@href}%
\providecommand \@@href[1]{\endgroup#1\@@endlink}%
\providecommand \@sanitize@url [0]{\catcode `\\12\catcode `\$12\catcode
  `\&12\catcode `\#12\catcode `\^12\catcode `\_12\catcode `\%12\relax}%
\providecommand \@@startlink[1]{}%
\providecommand \@@endlink[0]{}%
\providecommand \url  [0]{\begingroup\@sanitize@url \@url }%
\providecommand \@url [1]{\endgroup\@href {#1}{\urlprefix }}%
\providecommand \urlprefix  [0]{URL }%
\providecommand \Eprint [0]{\href }%
\providecommand \doibase [0]{http://dx.doi.org/}%
\providecommand \selectlanguage [0]{\@gobble}%
\providecommand \bibinfo  [0]{\@secondoftwo}%
\providecommand \bibfield  [0]{\@secondoftwo}%
\providecommand \translation [1]{[#1]}%
\providecommand \BibitemOpen [0]{}%
\providecommand \bibitemStop [0]{}%
\providecommand \bibitemNoStop [0]{.\EOS\space}%
\providecommand \EOS [0]{\spacefactor3000\relax}%
\providecommand \BibitemShut  [1]{\csname bibitem#1\endcsname}%
\let\auto@bib@innerbib\@empty
\end{thebibliography}%


\begin{thebibliography}{}
%
%
\bibitem{Rapol2004}
U.~D.~Rapol, A.~Krishna, A.~Wasan and V.~Natarajan, The European Physical Journal D - Atomic, Molecular, Optical and Plasma Physics. \textbf{29}, (2004) 409-414.
\bibitem{Kostylev2014} 
N.~Kostylev, E.~Ivanov, M.~E.~Tobar and J.~J.~McFerran, J. Opt. Soc. Am. B \textbf{31}, \textit{7} (2014) 1614-1620.
\bibitem{Ender1993}
V.~Enders, PH.~Courteille, R.~Huesmann, L.~S.~Ma, W.~Neuhauser
R.~Blatt and P.~E.~Tosche, Europhys. Lett. \textbf{24}, \textit{5} (1993) 325-331.
\bibitem{Hansch2006}
T.~H\"{a}nsch, Rev. Mod. Phys. \textbf{78} (2006) 1297.
\bibitem{Cornell2002} 
E.~A.~Cornell and C.~E.~Wieman, Rev. Mod. Phys. \textbf{74}, \textit{3} (2002) 875-893.
\bibitem{Takasu2003} 
Y.~Takasu, K.~Maki, K.~Komori, T.~Takano, K.~Honda, M.~Kumakura, T.~Yabuzaki and Y.~Takahashi, Phys. Rev. Lett. \textbf{91}, \textit{4} (2003).
\bibitem{McLoughlin2011} 
J.~J.~McLoughlin, A.~H.~Nizamani, J.~D.~Siverns, R.~C.~Sterling, M.~D.~Hughes,
B.~Lekitsch, B.~Stein, S.~Weidt and W.~K.~Hensinger, Phys.~Rev.~A \textbf{83}, \textit{1} (2011) 013406.
\bibitem{Rohde2010}
F.~Rohde, M. Almendros, C. Schuck, J, Huwer, M. Hennrich and J. Eschner, J. Phys. B \textbf{43}, (2010) 115401.
\bibitem{Shao2019}
X.~Shao, H.~Han, Y.~Su, H.~Wang, Z.~Zhang, S.~Fang, G.~Chang and Z.~Wei, IEEE CLEO (2019) 1-2
\bibitem{Zhao1998} 
W.~Z.~Zhao, J.~E.~Simsarian, L.~A.~Orozco and G.~D.~Sprouse, Rev. Sci. Instrum. \textbf{69}, \textit{11} (1998) 3737-3740.
\bibitem{Mohle2013}
K.~M\"{o}hle, E.~V.~Kovalchuk, K.~D\"{o}ringshoff, M.~Nagel and A.~Peters, Appl. Phys. B  \textbf{111}, (2013) 223-231.
\bibitem{Saleh2015}
K.~Saleh, J.~Millo, A.~Didier, Y.~Kersal\'{e}, and C.~Lacro\^{u}te, Appl. Opt. \textbf{54} (2015) 9446.
\bibitem{Couturier2018}
L.~Couturier, I.~Nosske, F.~Hu , C.~Tan, C.~Qiao, Y.~H.~Jiang, P.~Chen and M.~Weidemüller, Rev. Sci. Instrum. \textbf{89}, (2018) 043103.
\bibitem{Hall2006}
J.~L.~Hall, Rev. Mod. Phys. \textbf{78}, (2006) 1279-1295.
\bibitem{Demtroeder}
W.~Demtr\"{o}der, \textit{Laser spectroscopy} (Springer, 2003).
\bibitem{Bridge2009}
E.~M.~Bridge, J.~Millen, C.~S.~Adams and M.~P.~A.~Jones, Rev. Sci. Instrum. \textbf{80}, (2009) 013101.
\bibitem{Jayakumar2015}
A.~Jayakumar, B.~Plotkin-Swing, A.~O.~Jamison and S.~Gupta, Rev. Sci. Instrum. \textbf{86}, (2015) 073115.
\bibitem{Skolnick}
M.~Skolnick, IEEE J. Quant. Elec. \textbf{6}, (1970) 139.
\bibitem{Kim2003}
J.~I.~Kim, C.~Y.~Park, J.~Y.~Yeom, E.~B.~Kim and T.~H.~Yoon, Opt. Lett. \textbf{28}, \textit{4} (2003) 245-247.
\bibitem{Nizamani} 
A.~H.~Nizamani, J.~J.~McLoughlin and W.~K.~. Hensinger, Phys. Rev. A \textbf{82}, (2010) 043408.
\bibitem{Wang2011} 
W.-L.~Wang, X.-Y.~Xu, Chin. Phys. Lett. \textbf{28}, (2011) 033202.


\bibitem{Streed2008} 
E.~W.~Streed, T.~J.~Weinhold and D.~Kielpinski, Appl. Phys. Lett. \textbf{93}, \textit{071103} (2008).
\bibitem{Zhu2014} 
S.~Zhu, T.~Chen,~X.~Li and Y.~ Wang, J. Opt. Soc. Am. B. \textbf{31}, (2014) 2302-2309. 
\bibitem{Biercuk2014} 
M.~W.~Lee, M.~C.~Jarratt, C.~Marciniak and M.~J.~Biercuk, Opt. Express \textbf{22}, (2014) 7210-7221.
\bibitem{Smeets2003}
B.~Smeets, R.~C.~M.~Bosch, P.~Van~Der~Straten, E.~Te~Sligte, R.~E.~Scholten, H.~C.~W.~Beijerinck and K.~A.~H.~Van~Leeuwen, Appl. Phys. B \textbf{76}, (2003) 815-819.
\bibitem{Kunz2013}
P.~D.~Kunz, T.~P.~Heavner and S.~R.~Jefferts, Appl. Opt. \textbf{52}, (2013) 8048-8053.
\bibitem{Haroche1972}
S.~Haroche and F.~Hartmann, Phys. Rev. A \textbf{6}, (1972) 1280–1300.
\bibitem{Park2003} 
C.~Y.~PARK and T.~H.~YOON, Jpn. J. Appl. Phys. \textbf{42}, \textit{7A} (2003) L754–L756.
\bibitem{Sargent}
P.~Meystre and M.~Sargent, \textit{Elements of Quantum Optics} (Springer, 2007)
\bibitem{Bjorklund1980} 
G.~C.~Bjorklund, Opt. Lett. \textbf{5}, (1980) 15–17.
\bibitem{Kobayashi1982} 
S.~Kobayashi, Y.~Yamamoto, M.~Ito and T. Kimura, IEEE J. Quantum Electron. \textbf{QE-18}, \textit{4} (1982) 582-595.
\bibitem{Lenth1983} 
W.~Lenth, Opt. Lett. \textbf{8}, \textit{11} (1983) 575-577.
\bibitem{Rana2012} 
R.~Rana, M.~H\"{o}cker and E.~G.~Myers, Phys. Rev. A \textbf{86}, \textit{05} (2012) 050502.
\bibitem{Laeter2006} 
J.~R.~de~Laeter and N.~Bukilic, Int. J. Mass Spectrom. \textbf{252}, \textit{3} (2006) 222-227.
\bibitem{Meija2016} 
J.~Meija, T.~B.~Coplen, M.~Berglund, W.~A.~Brand, P.~De~Bièvre, M.~Gr\"{o}ning, N.~E.~Holden, J.~Irrgeher, R.~D.~Loss, T.~Walczyk and T.~Prohaska, Pure Appl. Chem. \textbf{88}, \textit{3} (2016) 293–306.
\bibitem{Fisk1997} 
P.~T.~H.~Fisk, M.~J.~Sellars, M.~A.~Lawn and C.~Coles, IEEE Trans. Ultrasonics Ferroelectrics Frequency \textbf{44}, \textit{2} (1997) 344–354.
\bibitem{Blatt1983} 
R.~Blatt, H.~Schnatz and G.~Werth, Z Physik A \textbf{312}, \textit{3} (1983) 143-147.

\bibitem{Das2005} 
D.~Das, S.~Barthwal, A.~Banerjee and V.~Natarajan, Phys. Rev. A \textbf{72}, \textit{3} (2005) 032506.

\bibitem{wang2017} 
Y.~Wang, M.~Um, J.~Zhang, S.~An, M.~Lyu, J.-N~Zhang, L.-M. Duan, D.~Yum and K.~Kim, Nature Photonics \textit{11}, (2017) 646–650.
\bibitem{Olmschenk2007} 
S.~Olmschenk, K.~C.~Younge, D.~L.~Moehring, D.~N.~Matsukevich, P.~Maunz and C.~Monroe, Phys. Rev. A \textbf{76}, \textit{5} (2007) 052314.
\bibitem{Huntemann2016}
N.~Huntemann, C.~Sanner, B.~Lipphardt, Chr.~Tamm and E.~Peik, Phys. Rev. Lett. \textbf{116} (2016) 063001
\bibitem{Streed2012}
E.~W.~Streed, A.~Jechow, B.~G.~Norton and D.~Kielpinski, Nat. Commun. \textbf{3}, \textit{933} (2012).

\bibitem{Blums2017}
V.~Bl\={u}ms, M.~Piotrowski, M.~I.~Hussain, B.~G.~Norton, S.~C.~Connell, S.~Gensemer, M.~Lobino and E.~W.~Streed, Science Advances \textbf{4}, \textit{3} (2018) eaao4453.

\bibitem{Ghadimi2017} 
M.~Ghadimi, V.~Bl\={u}ms, B.~G.~Norton, P.~M.~Fisher, S.~C.~Connell, J.~M.~Amini, C.~Volin, H.~Hayden, C.-S.~Pai, D.~Kielpinski, M.~Lobino and E.~W.~Streed, npj Quantum Inf. \textbf{3}, \textit{4} (2017).
\bibitem{Hannaford1983} 
P.~Hannaford, Contemporary Physics \textbf{24}, \textit{3} (1983) 251-270.

\bibitem{Zhao2008} 
P.-Y.~Zhao,  Z.~-X.~Xiong, J.~Liang, L.~-X.~He and B.~-L.~Lu, Chin. Phys. Lett. \textbf{25}, \textit{10} (2008) 3631-3634.
\bibitem{Kimball1998}
D.~F.~Kimball, \textit{Collisional Perturbation of States in Atomic Ytterbium} (University of California at Berkeley, 1998)
\bibitem{Ghadimi2020} 
M.~Ghadimi, E.~M.~Bridge, J.~Scarabel, S.~Connell, K.~Shimizu, E.~Streed and M.~Lobino, \textit{Appl. Opt.} \textbf{59}, 5136-5141 (2020).
\end{thebibliography}

\section{Conclusions}
In summary we have successfully demonstrated the POLEAS locking technique with neutral Yb atoms in a hollow cathode discharge lamp for stabilising a $398.9$~nm laser to the $^1S_0$$\leftrightarrow$$^1P_1$ transition of several Yb isotopes. This locking technique has shown to provide the necessary frequency stability to reliably photoionise the selected isotope of ytterbium from a naturally abundant source. Furthermore, the frequency stability measured would be sufficient for laser decelerating Yb atoms in a Zeeman slower for experiments utilising neutral Yb. By using commercially available equipment and a simple optical setup our approach significantly simplifies that taken compared to designing and building a Yb vapour cell or transfer cavity. We present this as a simple, robust solution for frequency stabilising the $398.9$~nm laser for atomic physics experiments using Yb atoms and Yb$^{+}$ ions.

\section{Acknowledgements}

This research was financially supported by the Griffith University Research Infrastructure Programme, the Griffith Sciences equipment scheme, Australian Research Council Discovery  (DP130101613) and Linkage (LP180100096) Projects, ES, ML, and DK were supported by Australian Research Council Future Fellowships (FT130100472, FT180100055, FT110100513), VB, JS, KS, and SC were supported by the Australian Government Research Training Program Scholarship.

%

\end{document}